\documentclass[aps,prc,twocolumn,floatfix,nofootinbib,preprintnumbers,superscriptaddress,longbibliography]{revtex4-1}

\usepackage{epsfig}
\usepackage{graphicx}
\usepackage{stmaryrd}
\usepackage{amssymb,tabularx,dcolumn}
\usepackage{supertabular,ltxtable}
\usepackage{longtable}
\usepackage{amsmath}
\usepackage{amstext}
\usepackage{float}
\usepackage{color}
\usepackage{bm}
\usepackage{hyperref}
\hypersetup{breaklinks=true,colorlinks=true,linkcolor=blue,citecolor=blue,filecolor=magenta,urlcolor=cyan}
\usepackage{mathtools}
\usepackage{multirow}
\usepackage{makecell}
\usepackage[utf8]{inputenc}
\usepackage{tabu}
\usepackage{braket}
\usepackage{subfigure}
\usepackage{enumitem}

\makeatletter
\def\@bibdataout@aps{%
\immediate\write\@bibdataout{%
@CONTROL{%
apsrev41Control%
\longbibliography@sw{%
    ,author="08",editor="1",pages="1",title="0",year="1"%
    }{%
    ,author="08",editor="1",pages="1",title="",year="1"%
    }%
  }%
}%
\if@filesw \immediate \write \@auxout {\string \citation {apsrev41Control}}\fi
}
\makeatother

\renewcommand{\vec}{\boldsymbol}
\newcommand{\rf}{\langle r^4\rangle}

\begin{document}

\title{Nuclear charge densities in spherical and deformed nuclei: towards precise calculations of charge radii }

\author{Paul-Gerhard Reinhard}
\affiliation{Institut für Theoretische Physik, Universität Erlangen, Erlangen, Germany}

\author{Witold Nazarewicz}
\affiliation{Facility for Rare Isotope Beams, Michigan State University, East Lansing, Michigan 48824, USA}
\affiliation{Department of Physics and Astronomy, Michigan State University, East Lansing, Michigan 48824, USA}

\begin{abstract}
\begin{description}
\item[Background]
Precise measurements  of atomic transitions affected by electron-nucleus hyperfine interactions offer sensitivity  to explore  basic properties of the atomic  nucleus  and  study    fundamental  symmetries, including  the  search  for  new  physics  beyond  the  Standard Model  of  particle  physics. In particular, such measurements, augmented  by atomic and nuclear calculations, will  enable extraction of the higher-order radial moments of the charge density
distribution in spherical and deformed nuclei. The new data impose higher precision requirements on a theoretical description.

\item[Purpose]
The nuclear charge density is composed of the proton point distribution folded with the    nucleonic charge distributions. The latter induce subtle relativistic corrections due to the coupling of nucleon magnetic moments  with the nuclear
spin-orbit density. Additional corrections come from  the effect of center-of-mass projection. 
We assess the precision  of nuclear charge density calculations by studying the behavior of relativistic and  center-of-mass motion corrections  to the second and fourth charge radial moments. Special attention has been paid to  the magnetic spin-orbit density associated with the local  variations of the  spin-orbit current.

\item[Methods]
The calculations for semi-magic and open-shell nuclei are performed in the framework of self-consistent mean-field theory  using  quantified  energy density functionals and density-dependent pairing forces. We used the general expression
for the spin-orbit form factor that is valid for spherical and deformed nuclei.

\item[Results]
 We studied the impact of various correction terms on the charge radii, fourth radial moments, diffraction radii, and surface thickness  of spherical and deformed nuclei. The spin-orbit corrections to charge radial moments and surface thickness show strong shell fluctuations which can make an appreciable effect when aiming at high-precision predictions of isotopic shifts. The inclusion of relativistic and center-of-mass corrections  impacts the quality of  energy density functionals optimized to charge radii data.

\item[Conclusions]
To  establish reliable
constraints on the existence of new forces from isotope shift measurements, precise calculations of nuclear charge densities of deformed nuclei are needed. The proper inclusion of the spin-orbit charge density and other correction terms is essential when aiming at extraction of subtle effects which become particularly visible in isotopic trends. It is also important when developing high-quality nuclear energy density functionals optimized using heterogeneous datasets involving absolute charge radii, differential charge radii, and  charge form factor properties deduced from electron scattering data.

\end{description}
\end{abstract}

\maketitle

\section{Introduction}

High-precision studies of atomic transitions  offer   complementary information on the structure of  atomic nucleus and fundamental symmetries, including hints of new physics beyond the Standard Model of particle physics \cite{Saf18,Ber18,Sta18,Del17,Viatkina2019}.
In particular, precise measurements of transition frequencies  allow extraction of tiny variations in the root-mean-square (rms) nuclear charge radii across long isotopic chains of stable and radioactive nuclei \cite{Gar16,Cam16,Ham18,Mil19,Gor19,deGroote2020,Yordanov2020}. This carries the potential to constrain the existence of new forces and hypothetical particles with unprecedented sensitivity \cite{Gar16,Del17,Del17b,Fru17,Fla18,Mikami2017,Ber18,Sta18,Berengut2020}.
The theoretical findings have stimulated considerable developments in high-precision experimental techniques \cite{Gebert15,Bra19,Manovitz2019,Counts2020}.
The new unprecedented  level of precision offers sensitivity not only to explore new physics, but would also provide access to nuclear observables that have so far been elusive, such as  the fourth-order charge radial moment  $\rf$ \cite{Pap16,Fla18,Ekman2019} that carries information on nuclear surface properties \cite{Reinhard2020,Allehabi2020}.

In order to extract structural information from atomic measurements, it is important for nuclear theory to produce reliable predictions of nuclear charge densities and currents.
Nuclear models usually  yield  the proton and neutron densities from which  the nuclear charge density can be extracted by considering  several corrections  \cite{Friar1975,Friedrich1986}.
The spurious  center-of-mass  (c.m.) motion  is corrected by an unfolding with the width of the centre-of-mass vibrations.
The nucleon structure is taken into account  by folding with the intrinsic form factor of the free nucleons expressed in terms of the Sachs form factors. The leading part comes from the folding with the nucleonic charge form factors. Moreover, there are the magnetic form factors of
the nucleons which contribute to the charge density through the coupling to the nuclear spin-orbit density.  The latter contributions are called  the spin-orbit terms in the following.  Together with the relativistic Darwin-Foldy term, they  constitute the relativistic corrections to the charge density
\cite{deForest1966,Bertozzi1972,Friar1975,Friar1973,Martorell1980,Friar1997}. (For a recent discussion of the nucleonic corrections see also \cite{Lorce:2020onh}.)

The relativistic corrections are routinely considered in
few-body and many-body \emph{ab-initio} nuclear calculations of charge densities and related observables, see, e.g., Refs.~\cite{Hagen2016,Hoferichter2020}
In calculations of charge radii based on the self-consistent mean-field theory
\cite{Friedrich1986,Reinhard1989,Bender2003,Reinhard2013},
the proton  and neutron form factor can be expressed in terms of single-particle Hartree-Fock (HF) or Hartree-Fock-Bogoliubov (HFB) densities. The  spin-orbit correction is often neglected
\cite{Liang2018,Viatkina2019,Allehabi2020,Allehabi2020a} which is commonly considered reasonable for predictions that do not require high accuracy.
Moreover, the majority of calculations with relativistic corrections are
carried out in spherical geometry  \cite{Ong2010,Horowitz2012,Kurasawa2019,Kurasawa2020};  this becomes insufficient when considering  long isotopic chains which contain  deformed nuclei.

In this work, we take a fresh look at the calculation of nuclear radii under the perspective of the enhanced demands of precision as required by current measurements of nuclear radii. To that end, we apply the self-consistent mean-field theory to study nuclear charge densities and charge radial moments for spherical and deformed nuclei with emphasis on the  intrinsic nucleon form factors and the relativistic contributions  that are essential for the accuracy required for precision studies. The paper is organized as follows. The definitions of corrections to charge densities and charge radii given in Sec.~\ref{sec:Def}.
Section~\ref{sec:Framework} describes the theoretical approach used. This is followed by description of results in Sec.\ref{sec:Res}. Finally, Sec.~\ref{sec:Conclusions} contains the conclusions of our study.

\section{Key observables}\label{sec:Def}
\subsection{The charge form factor}

The nuclear charge density is uniquely related to the  nuclear charge form factor $F_c$:
\begin{equation}\label{eq:rhoc}
 \rho_c(\vec{r}) =\frac{1}{(2\pi)^3}\int d^3q 
    e^{-i\vec{q}\cdot\vec{r}}F_c(\vec{q}).
\end{equation}
The latter is the quantity measured by electron scattering experiments \cite{deForest1966} and used for conveniently including the folding by the intrinsic nucleon form factors and the c.m.  motion correction.
In the following, we recall the relativistic and non-relativistic expressions for $F_c$ and briefly discuss the treatment of the c.m. correction.

\subsubsection{The magnetic contribution to charge density in the relativistic mean-field
  theory}

The relativistic operator for the nuclear charge form factor $\hat{F}_c$ is
 the zeroth component of the charge current $\hat{J}_0$
and reads \cite{Friar1975}:
\begin{equation}
  \hat{F}_c(\vec{q})
  \equiv
  \hat{J}_0(\vec{q})
  =
  \sum_{t\in\{p,n\}}
  f_{1,t}(\vec{q})\hat{\gamma}_0
  -
  f_{2,t}(\vec{q})\frac{\hbar}{2mc}\hat{\vec{\alpha}}\!\cdot\!\vec{q},
\end{equation}
where $m$ is the nucleon mass, $\hat{\vec{\alpha}}$ is the three-vector of Dirac matrices
\cite{Bjo64aB}, $f_{1,t}(\vec{q})$ is the intrinsic nucleon charge form factor,
and $f_{2,t}(\vec{q})$ is the intrinsic nucleon magnetic form factor. The charge
form factor can be written as:
\begin{equation}
  F_c(\vec{q})
  =
  \sum_{t\in\{p,n\}}\left[
  f_{1,t}(\vec{q})F_t(\vec{q})
    -
  f_{2,t}(\vec{q})\frac{F_{\mathrm{tens},t}(\vec{q})\hbar}{2mc}\right],
\end{equation}
where the form factors
\begin{align}
\begin{aligned}
  F_t(\vec{q})
  &=
  \int d^3r\,e^{\mathbf{i}\vec{q}\cdot\vec{r}}\rho_t(\vec{r}),
\\
  F_{\mathrm{tens},t}(\vec{q})
  &=
  \int d^3r\,e^{\mathbf{i}\vec{q}\cdot\vec{r}}\rho_{\mathrm{tens},t}(\vec{r}),
  \end{aligned}
 \end{align} 
can be expressed in terms of relativistic densities
\begin{align}
\begin{aligned}
  \rho_t(\vec{r})
  &=
  \sum_{\alpha\in t}v_\alpha^2\overline{\psi}_\alpha\hat{\gamma}_0\psi_\alpha^{\mbox{}},
\\
  \rho_{\mathrm{tens},t}(\vec{r})
  &=
  -\mathrm{i}\vec{\nabla}\cdot
  \sum_{\alpha\in t}v_\alpha^2\overline{\psi}_\alpha
  \mathrm{i}\left(\begin{array}{cc}
    0 & \vec{\sigma} \\ \vec{\sigma} & 0
  \end{array}\right)
  \psi_\alpha^{\mbox{}},
\end{aligned}
\end{align}
where the four-component eigenstate of the Dirac equation $\psi_\alpha$ is the nucleonic single-particle (s.p.)
wave function, the $v_\alpha^2$ are the BCS or HFB canonical pairing occupations, and the  tensor  density $  \rho_{\mathrm{tens},t}$ together with the nucleon magnetic form factors yield the magnetic  contribution to the charge density.

The intrinsic nucleon form factors are usually expressed in terms of
the Sachs form factors  $G_E$ and $G_M$ as
\begin{align}
\begin{aligned}
  f_{1,t}(\vec{q})
  &=
  \frac{G_{E,t}(\vec{q})+{q}^2\mathcal{D}\mu_tG_{M,t}(\vec{q})}{1+{q}^2\mathcal{D}}
  \quad,
\\
  f_{2,t}(\vec{q})
  &=
  \frac{-G_{E,t}(\vec{q})+\mu_tG_{M,t}(\vec{q})}{1+{q}^2\mathcal{D}},
\end{aligned}
\end{align}
where
\begin{equation}
{\cal D}=\frac{\hbar^2}{(2mc)^2}
\end{equation}
and $\mu_{t}$ are the magnetic moments of the nucleon: $\mu_p=2.79$ and $\mu_n = -1.91$.

The above expressions for form factors do not depend on the geometry of the Dirac equation. 
 The explicit spherical-geometry expressions can be found in, e.g.,  Refs.~\cite{Horowitz2012,Kurasawa2019}.

\subsubsection{The magnetic contribution to charge density in non-relativistic mean-field theory}

The  expression for the form factor in non-relativistic models is obtained by  the expansion in powers of ${\cal D}\propto m^{-2}$ up to first order \cite{deForest1966}.  In this non-relativistic limit, the charge form factor reads \cite{Friar1975}:
\begin{align}\label{allFF}
F_{c}(\vec{q}) &= 
\sum_{t\in\{p,n\}}
 \Big[G_{E,t}(\vec{q})
   \left(1-{\textstyle\frac{1}{2}}\vec{q}^2\mathcal{D}\right)
   F_t(\vec{q})
\nonumber \\
  &\quad
  -  \mathcal{D}\left[2\mu_t G_M(\vec{q})-G_{E,t}(\vec{q})\right]
  F_{\ell s,t}(\vec{q})\Big].
\end{align}
 The form factors
\begin{align}\label{eq:Ftls}
\begin{aligned}
  F_{t}(\vec{q})
  &=
  \int d^3r\,
  e^{{i}\vec{q}\cdot\vec{r}}\rho_t(\vec{r}),
\\
  F_{{\ell s},t}(\vec{q})
  &=
  \int d^3r\,
  e^{i\vec{q}\cdot\vec{r}}\bm{\nabla}\!\cdot\!\vec{J}_t(\vec{r})
  \end{aligned}
 \end{align}
are given in terms of the local  particle densities $\rho_t(\vec{r})$ and  spin-orbit currents
$\vec{J}_t(\vec{r})$: 
\begin{align}
\begin{aligned}
  \rho_t(\vec{r})
  &=
  \sum_\alpha v_\alpha^2 |\varphi_{t\alpha}(\vec{r})|^2,
\\
  \vec{J}_t(\vec{r})
  &=
i\sum_{\alpha}v_\alpha^2 {\varphi}_{t\alpha}^*(\vec{r})
   (\bm{\sigma}\!\times\!\bm{\nabla}){\varphi}_{t\alpha}^{\mbox{}}(\vec{r}),
  \end{aligned} 
 \end{align}  
with ${\varphi}_{t\alpha}^{\mbox{}}(\vec{r})$ being the canonical HFB (or BCS) wave functions
and  $v_\alpha^2$  the corresponding pairing occupation coefficients.
Note that the above derivation does not make assumptions about spatial
symmetries. Consequently, the expressions can be applied in  3D HFB codes as well as in 2D axial
or 1D spherical HFB calculations.

There is a subtle difference in the interpretation of the relativistic and  non-relativistic expressions.  In the relativistic form factor, the magnetic contributions are associated simply with the tensor density. 
In the non-relativistic case, this becomes the spin-orbit density and it turns out to be of the same order ${\cal D}$ as the relativistic Darwin term $G_{E}\mathcal{D}$. It is customary, to consider the purely electric contribution $G_{E,t}F_t$ as the leading term and everything else as relativistic correction.

\subsubsection{The center-of-mass contribution to charge density}

There are several ways to describe the c.m.
correction formally \cite{Schmid1991,Mihaila1999,Hagen2009}. For calculations in coordinate-space basis, as in this work, the most appropriate is the approximate projection technique \cite{Schmid1991}. 
 In the second-order Gaussian overlap approximation, the projected
point-proton form factor can be written as:
\begin{equation}\label{com}
F_{p}^{\rm proj}(\vec{q})=    F_{p}(\vec{q})   \exp\left(\frac{3}{8\langle\Phi|\hat{\vec{P}}^2|\Phi\rangle}
     \vec{q}^2\right),
\end{equation}
where $|\Phi\rangle$ is the BCS or HFB state and $\vec{P}$ is the c.m. momentum. This  expression is analogous to that obtained in the 
harmonic oscillator expansion  \cite{Mihaila1999,Hagen2009}.

The quality of the approximation (\ref{com}) has been examined in
Ref.~\cite{Schmid1991}. It was concluded that the bulk nuclear 
properties as the diffraction radius and surface thickness are more
robust and show little difference between approximate and exact
projection. Noteworthy effects appear only for light nuclei up to Ca,
reaching to typically 0.0004\,fm uncertainty for Ca. 

\subsection{The charge  radius}

The squared charge radius is obtained from the charge form factor $F_{c}(\vec{q})$ as
\begin{equation}
  \langle r_{c}^2\rangle 
  =
  -\frac{\bm{\nabla}^2F_c(\mathrm{q})\Big|_{q=0}}{F_c(0)}
  \quad.
\end{equation}
For the reflection-symmetric nuclei, all form factors $F(\mathrm{q})$ in Eq.~(\ref{allFF})  fulfill the condition:
$\vec{\nabla} F(\mathrm{q})=0$. The product rule with ${\nabla}^2$ then
yields only terms with zeroth or second derivative. We abbreviate
${\nabla}^2f|_{q=0}=f''$ for each factor in the form factor and insert
the values in zeroth order $G_{E,p}(0)=1$, $G_{E,n}(0)=0$, $G_M(0)=1$,
$F_p(0)=Z$, $F_n(0)=N$, and $F_{{\ell s},t}(q)(0)=0$.  This yields at $q=0$:
\begin{flalign}
  F_c
  &=
  Z, \\
  F''_c
  &=
  F''_p+ZG''_{E,p}-3Z\mathcal{D} + NG''_{E,n} 
 \nonumber \\&  -
  (2\mu_p-1)\mathcal{D}F''_{\ell s,p}
  -
  2\mu_n\mathcal{D}F''_{{\ell s},n}.
\end{flalign}
The second derivatives can be written
as
\begin{flalign}
  F''_p
  &=
  \int d^3r\,{r}^2\rho_p(\vec{r})
  \equiv
  Z\langle r^2\rangle_{pp},
\\
  F''_{\ell s,t}
  &=  \int d^3r\,{r}^2\vec{\nabla}\cdot\vec{J}_t(\vec{r}) \label{fpp}
\end{flalign}
and similarly $G''_{E,p}=\langle r^2\rangle_p^\mathrm{(intr)}$,
 $G''_{E,n}=\langle r^2\rangle_n^\mathrm{(intr)}$.
 In the above expression, $\langle r^2\rangle_{pp}$ indicates the point-proton radius as it emerges directly from the mean-field calculation. The quantity $ F''_{\ell s,t}$ can alternatively be written as:
 \begin{equation}\label{fppls}
    F''_{\ell s,t} = 
  -2\langle\hat{\bm{\sigma}}\!\cdot\!\hat{\vec{\ell}}\rangle_t.
\end{equation} 
which adds a physical interpretation. In practice, we evaluate $F''_{\ell s,t}$ in terms of Eq. (\ref{fpp}) because the local spin-orbit current $\vec{J}_t$ is already provided by the HFB calculations.

By combining all contributions, we obtain the expression or the average squared charge radius:
\begin{equation}\label{eq:r2}
  \langle r^2_c\rangle
  =
  \langle r^2_{pp}\rangle
  +
  \langle r^2_p\rangle^\mathrm{(intr)}
  +
  \frac{N}{Z}\langle r^2_n\rangle^\mathrm{(intr)} +   \langle r^2\rangle^\mathrm{(rel)},
  \end{equation}
 where
 \begin{equation}\label{eq:rrel}
\langle r^2\rangle^\mathrm{(rel)}=
  3{\cal D} 
  +(\mu_p-{\textstyle\frac{1}{2}})\frac{4}{Z}{\cal D}
   \langle\hat{\vec{\sigma}}\!\cdot\!\hat{\vec{\ell}}\rangle_p
  +\mu_n\frac{4}{Z}{\cal D}
   \langle\hat{\vec{\sigma}}\!\cdot\!\hat{\vec{\ell}}\rangle_n
\end{equation}
is the relativistic contribution to the charge radius. As discussed above, it consists of the Darwin-Foldy (DF) term
$3{\cal D}$ and  the spin-orbit corrections. 

 The  form of
the spin-orbit terms in Eq.~(\ref{fpp}) that involves
$\vec{\nabla}\cdot\vec{J}$ is valid for arbitrary mean-field geometry. The second form (\ref{fppls}), involving
$  \langle\hat{\vec{\sigma}}\!\cdot\!\hat{\vec{\ell}}\rangle_t$,
is particularly useful if the spherical geometry is imposed. In this case,
the expectation value of the spin-orbit term  becomes independent of the radial profile of the wave functions and the expression reduces (for each nucleon type) to 
$\langle\hat{\vec{\sigma}}\!\cdot\!\hat{\vec{\ell}}\rangle=\sum_\alpha v_\alpha^2(\sigma \ell)_\alpha$ where  $(\sigma l)_\alpha=j_\alpha(j_\alpha+1)-l_\alpha(l_\alpha+1)-\frac{3}{4}$, which is $\ell_\alpha$ for  $j_\alpha=\ell_\alpha+1/2$ and $-(\ell_\alpha+1)$ for $j=\ell_\alpha-1/2$.
It is immediately seen that if both sub-shells of the spin-orbit doublet are occupied with the same weight, their  contribution to the $\ell s$ term in (\ref{eq:r2}) vanishes (spin-saturated case). The maximal spin-orbit contribution is attained when the lower-energy member of the  spin-orbit doublet is fully occupied and the upper-energy member with $j=\ell-1/2$ is not \cite{Friar1975}.

\section{Computational Framework}\label{sec:Framework}

The examples presented here were computed with non-relativistic nuclear density-functional theory (DFT) using the well known Skyrme energy-density functional, for a detailed review see \cite{Bender2003}. 
In our applications, we employ the Skyrme parametrization SV-bas from Ref.~\cite{Klu09a}
which has been optimized to a large experimental calibration dataset including information on several exotic nuclei. This is appropriate for the present study, which covers long isotopic and isotonic chains. 
 We have repeated calculations presented in this work  with   other Skyrme parametrizations and obtained results that are very similar to those with SV-bas. We also employed  the Fayans functional Fy($\Delta r$, HFB)
\cite{Rei17a,Mil19}, which uses the optimization dataset of SV-bas and adds to it a crucial new input consisting of differential  charge radii in the calcium chain.

To cover  deformed nuclei, we use the recently published code \texttt{SkyAx} which allows for  deformed axially symmetric shapes \cite{Rei21aR}. A word is in order about the treatment of pairing. The code \texttt{SkyAx} implements pairing at the BCS level 
using a soft cutoff in pairing space with the cutoff profile
as used in Ref.~\cite{Kri90a}
\begin{equation}
  w_\alpha
  =
  \left[1+
    \exp{\left((\varepsilon_\alpha-(\epsilon_{\mathrm{F},q_\alpha}+\epsilon_\mathrm{cut}))
            /\Delta\epsilon\right)}
  \right]^{-1}
\label{eq:softcut}
\end{equation}
where $\varepsilon_\alpha$ are the s.p. energies, 
 $\epsilon_\mathrm{cut}$ marks the cutoff band, and $\Delta\epsilon=\epsilon_\mathrm{cut}/10$ is its width.
We use a dynamical 
setting of the pairing band where $\epsilon_\mathrm{cut}$ is adjusted such
that a fixed number of nucleons $N_q+\eta_\mathrm{cut}N_q^{2/3}$ is
included in the sum $\sum_{\alpha\in q}w_\alpha$ \cite{Ben00a}, here
with $\eta_\mathrm{cut}=1.65$ for SV-bas (as in Ref.~\cite{Klu09a}) and $\eta_\mathrm{cut}=5$ for Fy($\Delta r$,HFB) as in \cite{Rei17a}.

It is to be noted that mere BCS is not always appropriate for nuclei at the edges of stability \cite{Dobaczewski1984,Dobaczewski1996,Mil19}, for which one should use, in principle, the full HFB framework. In this study, however,
we limit  the selection generally to nuclei whose proton and neutron Fermi energies are sufficiently bound  so the unphysical particle gas effects are avoided. 

The intrinsic form factors of the nucleons
were computed as in Ref.~\cite{Reinhard2013} with the Sachs form factors 
taken from Refs.~\cite{Simon1980,Walther1986}.
We wish to emphasize that
we do not use Eq.~(\ref{eq:r2}) to estimate charge radii  but rather compute numerically the radial moments (as well as  other observables directly from the charge density and charge form factor)  by the  folding the point charge distribution with the intrinsic nucleon form factors. In this way, all contributions to the charge density are automatically included.
In this work, we consider  subtle effects stemming from the relativistic corrections that place great demands on the accuracy of underlying calculations. In order to compute charge radii with precision better than 0.001\,fm, the calculations were carried out with enhanced demands on grid spacing, box size, Fourier transform, and HF+BCS termination criteria. 

\begin{figure}[htb]
\includegraphics[width=0.8\linewidth]{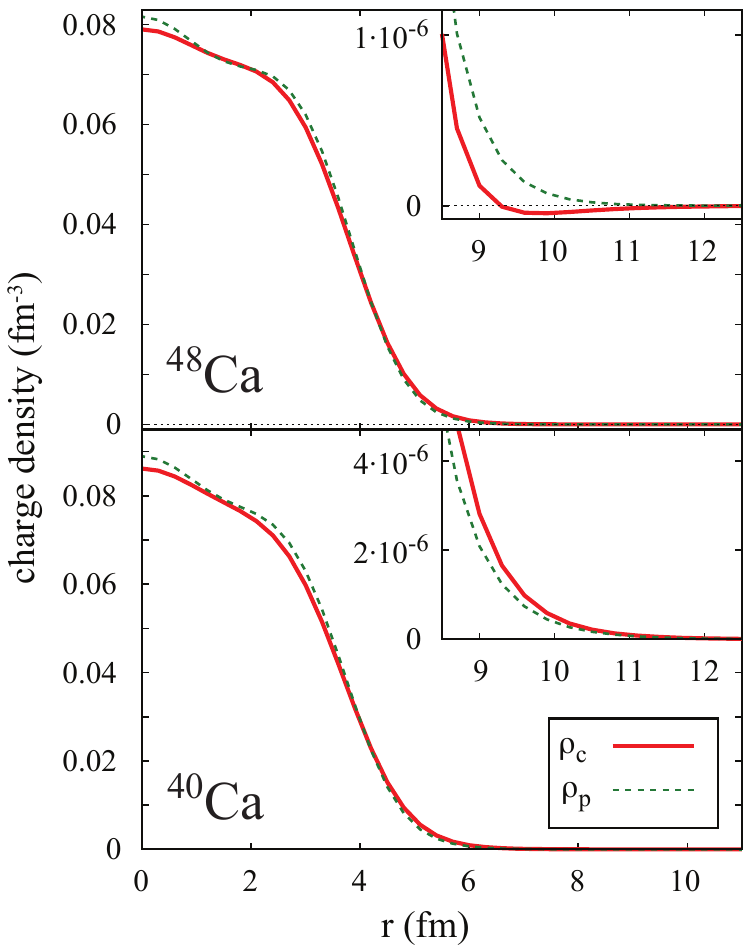}
\caption{Comparison of charge and proton densities for $^{40}$Ca  and $^{48}$Ca computed with SV-bas. The density dependence at large distances is shown in the insets.}
\label{fig:densities}
\end{figure}

\section{Results}\label{sec:Res}

We shall begin  from a pedagogical Fig.~\ref{fig:densities} showing the  charge density (\ref{eq:rhoc})  predicted with SV-bas
for $^{40}$Ca  and $^{48}$Ca. 
It is seen that at large distances the  neutron charge distribution and, to a lesser extent, the neutron spin-orbit density produce a negative contribution to the charge density in $^{48}$Ca,
while the effect of correction terms to the proton density in $^{40}$Ca is less pronounced.
The resulting negative contribution to the charge radius
helps bringing the charge radius of $^{48}$Ca very close to the value in $^{40}$Ca
\cite{Emrich1983,Horowitz2012,Hagen2016,Gar16,Kurasawa2019}, see discussion around Table~\ref{tablefit} below..

\begin{figure}[htb]
\includegraphics[width=\linewidth]{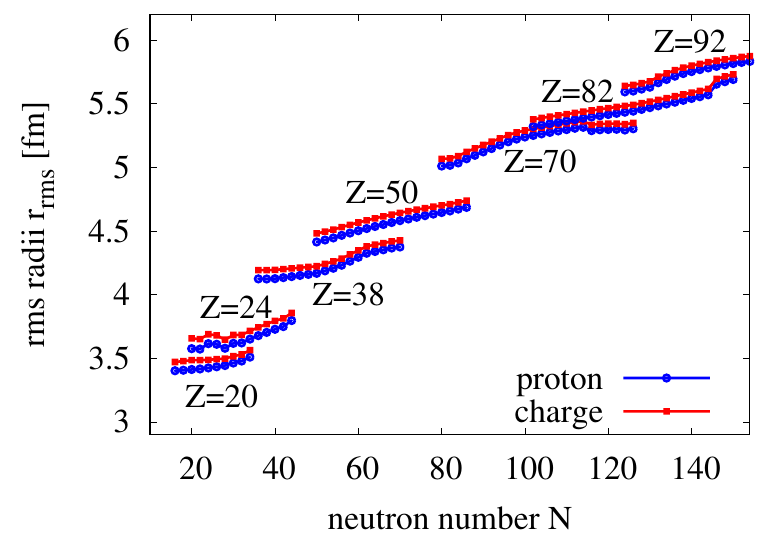}
\caption{\label{fig:rmsisotopic}
The rms point-proton (blue) and charge radii (red) for  isotopic chains of magic (Ca, Sn, Pb) and open-shell
(Cr, Sr, Yb, U) nuclei computed with SV-bas.
}
\end{figure}

Figure~\ref{fig:rmsisotopic} shows the predicted  rms proton and charge radii along  selected isotopic 
 chains which cover spherical and deformed nuclei. 
 The nucleonic and relativistic corrections are of the order of 0.05\,fm. This suggests that in applications where one aims merely at a global description of radii one may use the approximate relation \cite{Friar1975}
 \begin{equation}\label{eq:approxr}
\langle r^2_c\rangle
 \approx 
  \langle r^2_{pp}\rangle
  +
\langle r^2_p\rangle^\mathrm{(intr)} +
(N/Z) \langle r^2_n\rangle^\mathrm{(intr)}
\end{equation} 
with the constant proton and neutron  charge radii:
$ \langle r_p\rangle^\mathrm{(intr)}=0.848$\,fm \cite{Grinin2020} and
$\langle r^2_n\rangle^\mathrm{(intr)}=-0.1161$\,fm$^2$ \cite{PDG},
which is the radius correction (\ref{eq:r2}) without the relativistic term $\langle r^2\rangle^\mathrm{(rel)}$. (Note that the previous implementation of the proton form factor in Refs.~\cite{Simon1980,Walther1986,Reinhard2013}
implies the older value of the proton radius $\langle r_p\rangle^\mathrm{(intr)}=0.854$\,fm
which amounts to a constant reduction of about 0.001 fm, with  no  effect on trends.)

\begin{figure}[htb]
\includegraphics[width=1.0\linewidth]{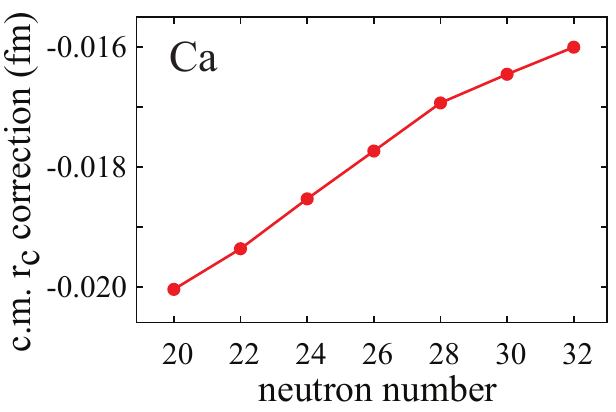}
\caption{\label{fig:cmcorr}
The c.m. correction to the charge radii of  Ca isotopes computed with SV-min.
}
\end{figure}

Figure\,\ref{fig:cmcorr} shows the c.m. correction to the charge radii of  Ca isotopes. It is seen that
the c.m. correction varies very smoothly with neutron number. Such a smooth trend holds also for the possible systematic error from approximate c.m. projection. Consequently, this already small error becomes reduced for differential radii.
We can thus conclude that small errors on
the charge radii due to the c.m. treatment have  negligible consequences for differential
radii,  for which  high precision is required.

On the other hand, the relativistic correction must be included in precision calculations  (which aims at average uncertainties as low as 0.015 fm) and studies of
small local variations of charge radii  such as the discontinuities across shell closures, which requires accuracy on charge radius prediction well below 0.01\,fm \cite{Gor19}.
\begin{table}
\caption{\label{tablefit} Properties of charge density computed with and without
spin-orbit contribution to the charge form factor for two 
energy density functionals: SV-bas and  Fy($\Delta r$,HFB). The rms  deviations from data for the diffraction radius, surface thickness, and charge radius (all in $10^{-3}$\,fm) are $\Delta R_c$, $\Delta \sigma_c$, and
$\Delta r_c$, respectively. 
We compute $R_c$ and $\sigma_c$ consistently from the charge form factor $F_c$ as discussed in  \cite{Rei21aR}.
The differential mean-square  charge radii for the Ca isotopes are defined in the usual way:
$\delta\langle r^2\rangle^{A',A}=
\langle r^2\rangle(^A{\rm Ca})-\langle r^2\rangle(^{A'}{\rm Ca})$. Their experimental values are $\delta\langle r^2\rangle^{40,48}=$0.007\,fm$^2$ and
$\delta\langle r^2\rangle^{44,48}=$0.308\,fm$^2$
The $\chi^2$ is the overall quality measure  for the fit, see \cite{Klu09a}. 
}
\begin{ruledtabular}
\begin{tabular}{lcccc}
 &  \multicolumn{2}{c} {SV-bas} & \multicolumn{2}{c}{Fy($\Delta r$,HFB)}
 \\
& full 
& no $\langle\hat{\vec{\sigma}}\!\cdot\!\hat{\vec{\ell}}\rangle_t$
& full
& no $\langle\hat{\vec{\sigma}}\!\cdot\!\hat{\vec{\ell}}\rangle_t$ \\[3pt]
\hline \\[-3pt]
$\Delta{R_c}$ & 34 & 33 & 29 & 27 \\
$\Delta\sigma_c$ & 26 & 29 & 17 & 20 \\
$\Delta r_c$ & 13 & 15 & 17 & 16 \\
$\delta\langle r^2\rangle^{40,48}$  & 0.109 &  0.205 & 0.010 & 0.112 \\
$\delta\langle r^2\rangle^{44,48}$  & $-$0.083 &  $-$0.128 & 0.294 & 0.235 \\[3pt]
$\chi^2$     & 56.9 & 62.6 & 65.1 & 125.8 \\
\end{tabular}
\end{ruledtabular} 
\end{table}
To quantify this point,  Table~\ref{tablefit} shows the impact of the spin-orbit contribution to the charge form factor on the results of the nuclear energy functional parametrizations SV-bas and Fy($\Delta r$,HFB) optimized to  large experimental datasets including form-factor information. 
The  r.m.s. deviations from experiment for the robust global observables, namely the
diffraction radius, surface thickness, and charge radius, depend, at first glance, weakly on  the spin-orbit correction. But note that the changes amount to 3--15\%,  which has a visible  impact on the overall
quality of the fit. This is already the case   for SV-bas, for which isotopic shifts have not  not included in the optimization dataset.
The differential mean-square  charge radii 
$\delta\langle r^2\rangle^{A',A}$ 
are more refined observables and they react quite dramatically. 
This becomes apparent in the huge change of $\chi^2$ for Fy(Dr,HFB) because this parametrization includes
the isotopic shifts in the fit data.
We conlcude that a description of isotope shifts without the spin-orbit correction is grossly misleading.

\begin{figure}[htb]
\includegraphics[width=0.9\linewidth]{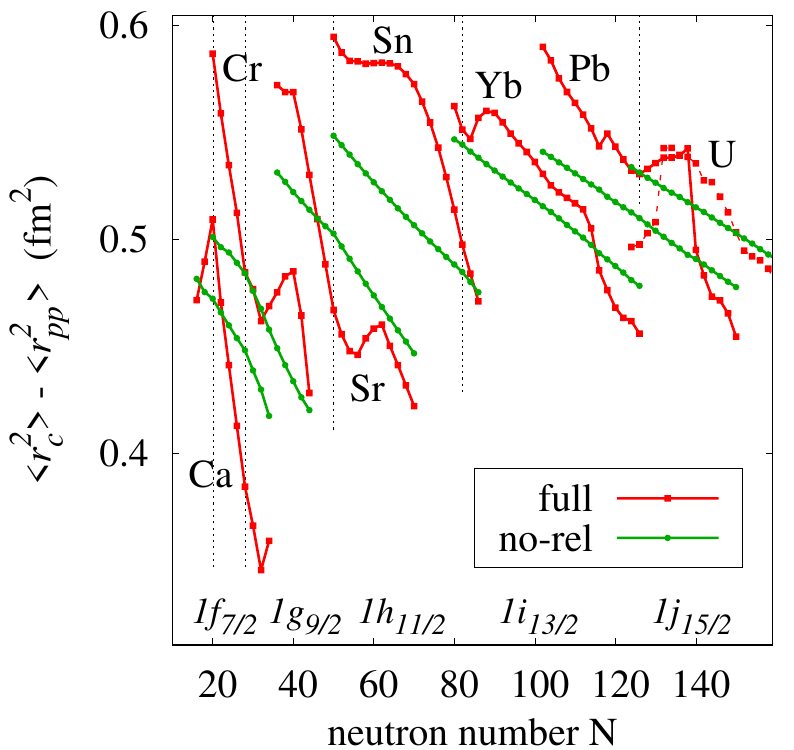}
\caption{The difference  $\langle r^2_c\rangle -\langle r^2_{pp}\rangle$ for several isotopic chains. For comparison, the results  without relativistic contribution (\ref{eq:rrel}) are shown.  Magic  numbers are indicated by vertical dashed lines. Positions of unique-parity shells are marked.}
\label{fig:trendsN}
\end{figure}

To show  the effect of spin-obit correction on charge radii in detail, Fig.~\ref{fig:trendsN} displays the difference  $\langle r^2_c\rangle -\langle r^2_{pp}\rangle$ for the isotopic chains of Fig.~\ref{fig:rmsisotopic}. For each chain, the results without the spin-orbit term exhibit a smooth decrease with neutron number that is consistent with the behavior of the intrinsic neutron charge distribution term in Eq.~(\ref{eq:approxr}).
As expected, the  spin-orbit contribution strongly fluctuates with $N$. In the regions corresponding to the gradual occupation of high-$j$ unique-parity shells the spin-orbit correction
rapidly decreases due to the negative value of $\mu_n$. The local increasing trends can be associated with the gradual occupation of the upper spin-orbit partner.  

The most dramatic local variation  of relativistic contributions is predicted between $^{40}$Ca and $^{52}$Ca (due to the population of 1$f_{7/2}$ and 2$p_{3/2}$ neutron shells)
and in the Sr chain around $N=50$  (due to the population of 1$g_{9/2}$ and 2$d_{5/2}$ neutron shells). In heavy nuclei the variations tend to be more gradual due to the fragmentation of the spin-orbit strength and the smoothing effect of pairing.

\begin{figure}[htb]
\includegraphics[width=0.9\linewidth]{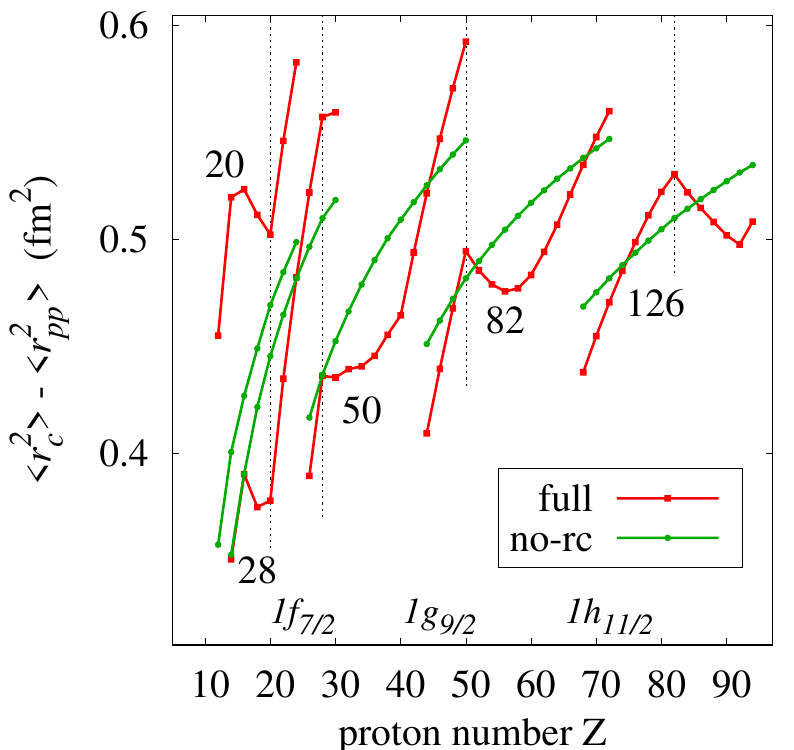}
\caption{Similar to Fig.~\ref{fig:trendsN} but for several isotonic chains of semi-magic nuclei.}
\label{fig:trendsZ}
\end{figure}

Figure~\ref{fig:trendsZ} illustrates the behavior of $\langle r^2_c\rangle -\langle r^2_{pp}\rangle$ along the isotonic chains
of semi-magic nuclei. Here, due to the positive value of $(\mu_p-1/2)$,  the spin-orbit contribution increases with $Z$ in the regions in which  high-$j$ shell are occupied.
The largest shell effect is predicted for $N=28$; it is see in the rapid rise of the spin-orbit correction between $^{48}$Ca and 
$^{56}$Ni.
Appreciable kinks in $\langle r^2_c\rangle$ are expected at $Z=50$ and 82 where the $j=\ell+1/2$ shells close up  and   the $j=\ell-1/2$ shells become occupied.

\begin{figure}[htb]
\includegraphics[width=0.8\linewidth]{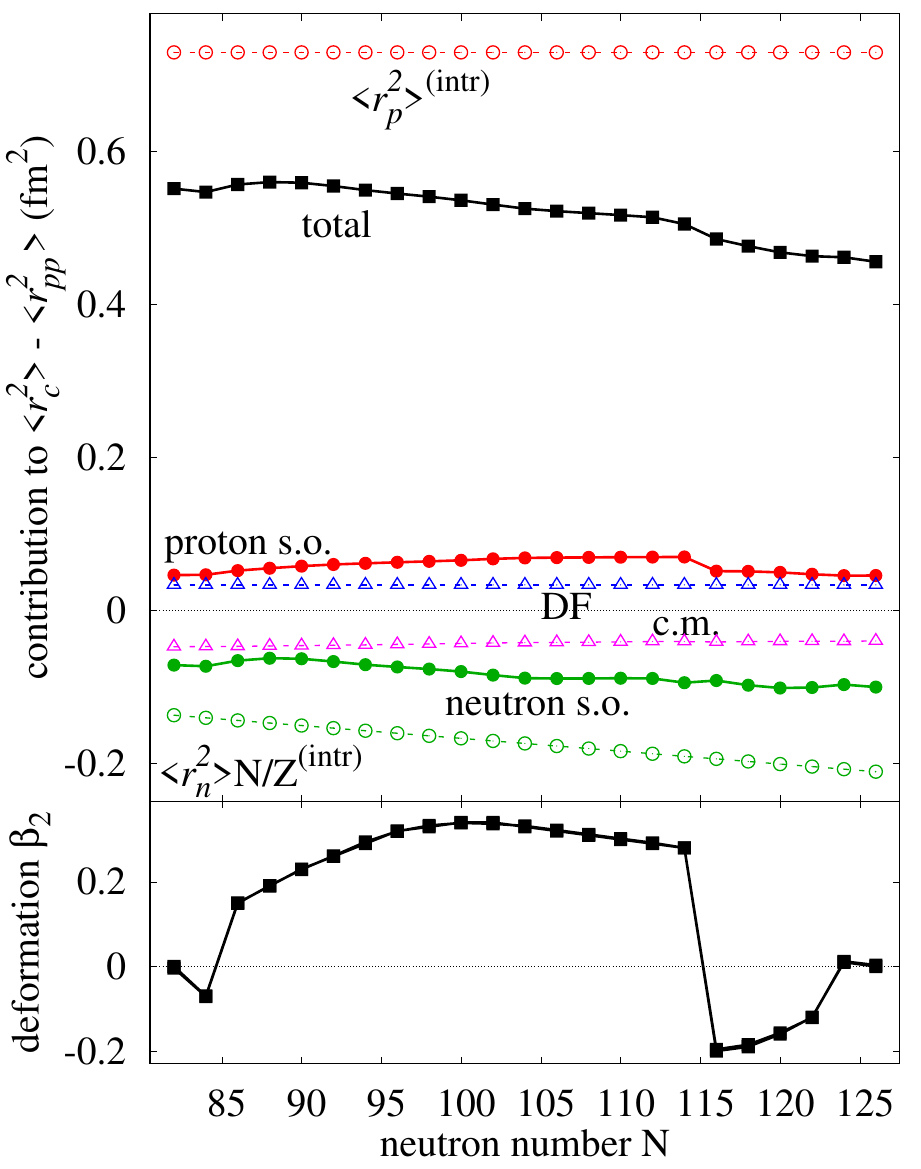}
\caption{Top: various corrections to $\langle r^2_c\rangle -\langle r^2_{pp}\rangle$ along
the chain of Yb isotopes. Bottom:  dimensionless quadrupole
shape deformation parameter $\beta_2=4\pi\langle Q_{20}\rangle/(5AR^2)$, where $Q_{20}$ is the mass quadrupole moment and $R_c$ the diffraction radius.
}
\label{fig:termsYb}
\end{figure}

To illustrate the impact of deformation effects, Fig.~\ref{fig:termsYb} shows  the corrections to the difference $\langle r^2_c\rangle -\langle r^2_{pp}\rangle$ along the Yb chain.
We note that 
 the deformed Yb isotopes are of particular interest in the context of ongoing experimental searches of new physics
\cite{Counts2020}. 
The intrinsic proton contribution, DF, and c.m. terms do not vary with $N$. The intrinsic neutron contribution shows the trivial linear $N/Z$ dependence. 
Note that in the deformed region the spin-orbit contributions change  gradually as the single-particle spin-orbit strength becomes highly fragmented by deformation and pairing. The prolate-to oblate shape transitions seen in the extremely proton-rich and extremely neutron-rich isotopes result in noticeable variations of spin-orbit contributions.

\begin{figure}[htb]
\includegraphics[width=0.9\linewidth]{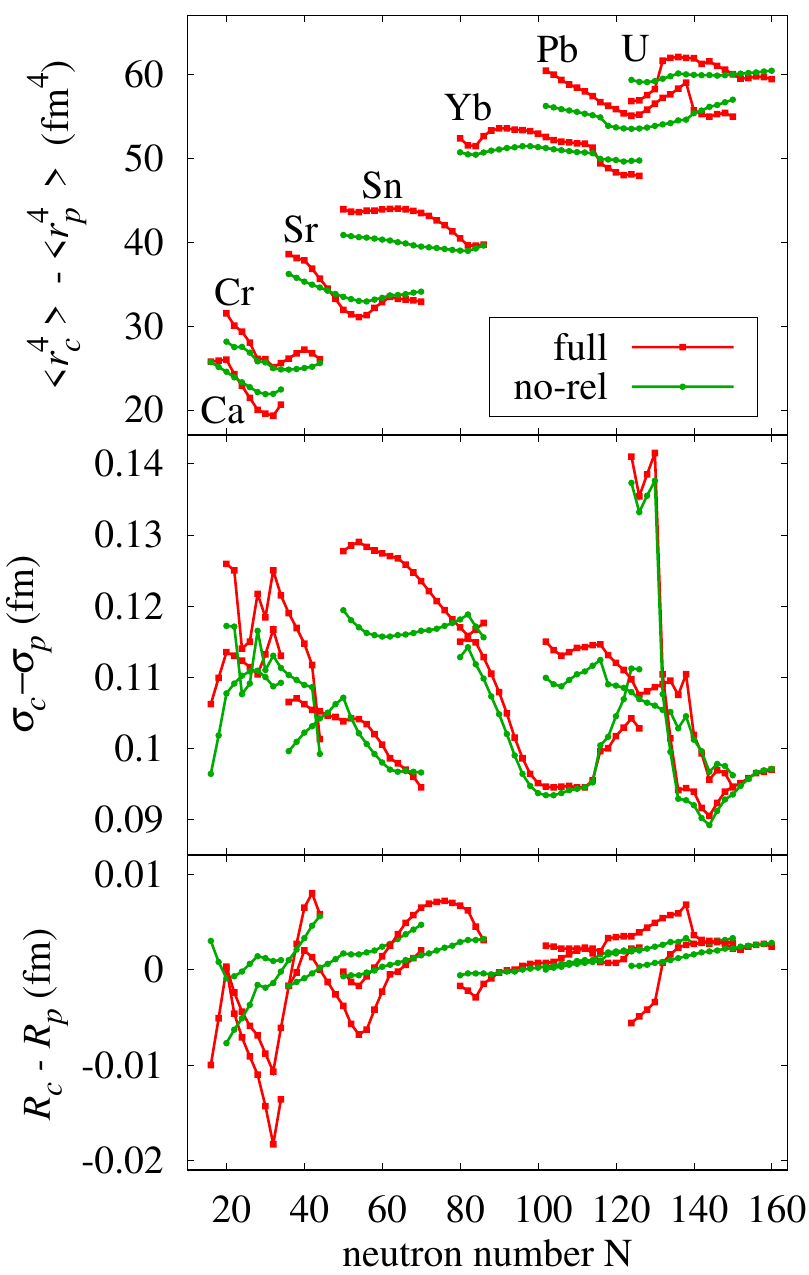}
\caption{The corrections to the
fourth radial  moment  $\rf$, surface thickness $\sigma$, and diffraction radius $R_c$ along several isotopic chains. For comparison, the results without relativistic contribution are also  shown.}
\label{fig:r4N}
\end{figure}

\begin{figure}[htb]
\includegraphics[width=0.9\linewidth]{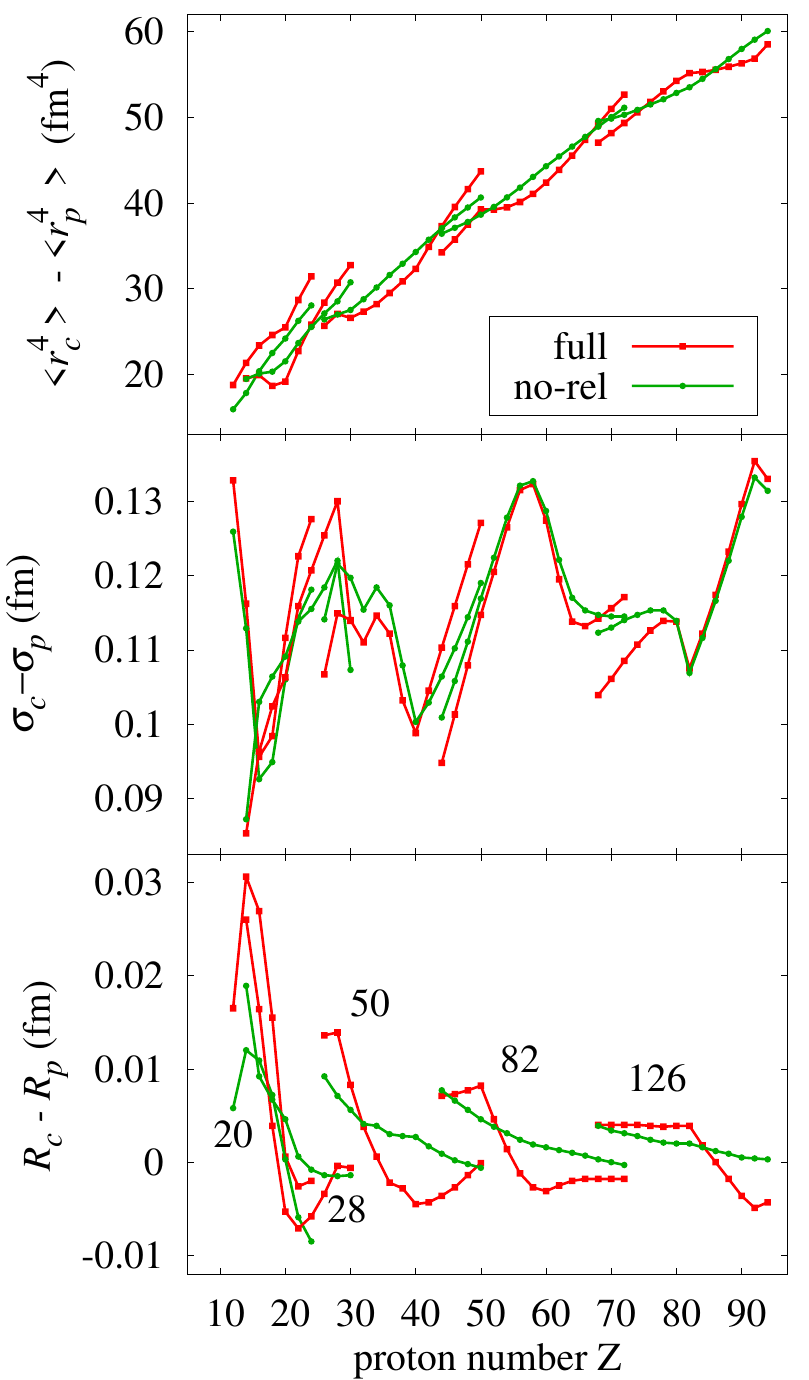}
\caption{Similar to Fig.~\ref{fig:r4N} but for
several  isotonic chains of semi-magic nuclei.}
\label{fig:r4Z}
\end{figure}

As demonstrated recently  \cite{Reinhard2020},
the fourth radial moment $\rf$ can be directly related to the surface
thickness $\sigma$ of nuclear density. (See also discussion in Ref.~\cite{Kurasawa2019}.) Precise knowledge of $\rf$ is essential to establish reliable
constraints on new physics.
 The fourth radial moment $\rf$ is computed from the charge density as obtained from $F_c$ by the inverse Fourier  transform (which we find the simplest and most robust procedure).
In order to demonstrate the sensitivity of $\langle r^4_c\rangle$  and bulk nuclear surface properties on the spin-orbit charge densities
Figs.~\ref{fig:r4N} and ~\ref{fig:r4Z} illustrate the impact of relativistic corrections  on $\rf$, surface thickness $\sigma$, and diffraction radii $R_c$ (see Ref.~\cite{Reinhard2020} for definitions).
It is seen that the shell fluctuations of relativistic corrections to these quantities are appreciable for $\rf$ and $\sigma_c$ while $R_c$ is less sensitive.

\section{Conclusions}\label{sec:Conclusions}
In this study, we investigated the impact of nucleonic corrections to the nuclear charge density and charge radial moments that are important in the context of precise  measurements  of  isotopic shifts.  The calculations were performed for spherical and deformed nuclei in the framework of self-consistent mean-field theory  using  quantified  nuclear energy density functionals and density-dependent pairing forces. We used the general expression
for the spin-orbit form factor that is valid for deformed nuclei.
The main conclusions and results of our study can be summarized
as follows:
\begin{enumerate}[label=(\roman*)]
\item
The nucleonic corrections are of the order of 0.05\,fm.  While the electric  nucleonic corrections to charge radii  do not depend on shell structure  and can be simply accounted for, the magnetic spin-orbit corrections strongly vary with particle number and require careful modelling. These  corrections can amount up to 0.01\,fm and need to be accounted for in precision studies  aiming at  extraction of tiny effects due to new physics from differential radii.

\item
Spin-orbit  corrections, with their pronounced shell effects, play a role  during the optimization of nuclear energy density functionals to the datasets involving charge radii.  On the other hand,
the uncertainty on
the charge radii due to the c.m. treatment has  negligible consequences for differential
radii, provided that the underlying EDF has been optimized to datasets including charge radii.

 \item 
 The discontinuities in charge radii across  shell closures
 results in kinks, which are  
 well below 0.01\,fm,  \cite{Gor19}. Since some of the nuclei of interest  are open-shell  systems \cite{Goodacre2020}, contributions from deformed spin-orbit densities can be appreciable.

\item
Deformation and pairing give rise to the fragmentation of the spin-orbit strength. This results in a smoothing of the spin-orbit correction to charge radii. To estimate this fragmentation for heavy nuclei, the deformed formalism laid out  in this work can be applied.

\item
It will be interesting to investigate experimentally the charge radii along the isotonic chains of semi-magic nuclei. Here, our calculations predict
a large shell effect  for $N=28$ that is characterized in the rapid rise of the spin-orbit correction between $^{48}$Ca and 
$^{56}$Ni. Also,
appreciable kinks in $\langle r^2_c\rangle$ are expected at $Z=50$ and 82 due to the closing of
proton $1g_{9/2}$ and  $1h_{11/2}$ intruder shells and filling the $1g_{7/2}$ and  $1h_{9/2}$ spin-orbit partner shells
\end{enumerate}

\begin{acknowledgements}
This material is based upon work supported by the U.S.\ Department of Energy, Office of Science, Office of Nuclear Physics under award numbers DE-SC0013365 and DE-SC0018083 (NUCLEI SciDAC-4 collaboration).
\end{acknowledgements}

\bibliography{radii}

\end{document}